\input harvmac
\input psfig
\input epsf
\noblackbox
\newcount\figno
 \figno=0
 \def\fig#1#2#3{
 \par\begingroup\parindent=0pt\leftskip=1cm\rightskip=1cm\parindent=0pt
 \baselineskip=11pt
 \global\advance\figno by 1
 \midinsert
 \epsfxsize=#3
 \centerline{\epsfbox{#2}}
 \vskip 12pt
 {\bf Fig.\ \the\figno: } #1\par
 \endinsert\endgroup\par
 }
 \def\figlabel#1{\xdef#1{\the\figno}}
 \def\encadremath#1{\vbox{\hrule\hbox{\vrule\kern8pt\vbox{\kern8pt
 \hbox{$\displaystyle #1$}\kern8pt}
 \kern8pt\vrule}\hrule}}
 %
 %


 \font\cmss=cmss10
 \font\cmsss=cmss10 at 7pt
 \def\rlx{\relax\leavevmode}
 \def\inbar{\vrule height1.5ex width.4pt depth0pt}
 \def\IC{\relax\,\hbox{$\inbar\kern-.3em{\rm C}$}}
 \def\IN{\relax{\rm I\kern-.18em N}}
 \def\IP{\relax{\rm I\kern-.18em P}}
 \def\ZZ{\rlx\leavevmode\ifmmode\mathchoice{\hbox{\cmss Z\kern-.4em Z}}
  {\hbox{\cmss Z\kern-.4em Z}}{\lower.9pt\hbox{\cmsss Z\kern-.36em Z}}
  {\lower1.2pt\hbox{\cmsss Z\kern-.36em Z}}\else{\cmss Z\kern-.4em
  Z}\fi}
 \def\IZ{\relax\ifmmode\mathchoice
 {\hbox{\cmss Z\kern-.4em Z}}{\hbox{\cmss Z\kern-.4em Z}}
 {\lower.9pt\hbox{\cmsss Z\kern-.4em Z}}
 {\lower1.2pt\hbox{\cmsss Z\kern-.4em Z}}\else{\cmss Z\kern-.4em
 Z}\fi}
 \def\IZ{\relax\ifmmode\mathchoice
 {\hbox{\cmss Z\kern-.4em Z}}{\hbox{\cmss Z\kern-.4em Z}}
 {\lower.9pt\hbox{\cmsss Z\kern-.4em Z}}
 {\lower1.2pt\hbox{\cmsss Z\kern-.4em Z}}\else{\cmss Z\kern-.4em Z}\fi}

 \def\narrowplus{\kern -.04truein + \kern -.03truein}
 \def\narrowminus{- \kern -.04truein}
 \def\narrowminussub{\kern -.02truein - \kern -.01truein}

 \def\frac#1#2{{#1\over #2}}

 \def\IZ{\relax\ifmmode\mathchoice
 {\hbox{\cmss Z\kern-.4em Z}}{\hbox{\cmss Z\kern-.4em Z}}
 {\lower.9pt\hbox{\cmsss Z\kern-.4em Z}}
 {\lower1.2pt\hbox{\cmsss Z\kern-.4em Z}}\else{\cmss Z\kern-.4em Z}\fi}
 \def\IB{\relax{\rm I\kern-.18em B}}
 \def\IC{{\relax\hbox{$\inbar\kern-.3em{\rm C}$}}}
 \def\Ic{{\relax\hbox{$\inbar\kern-.22em{\rm c}$}}}
 \def\ID{\relax{\rm I\kern-.18em D}}
 \def\IE{\relax{\rm I\kern-.18em E}}
 \def\IF{\relax{\rm I\kern-.18em F}}
 \def\IG{\relax\hbox{$\inbar\kern-.3em{\rm G}$}}
 \def\IGa{\relax\hbox{${\rm I}\kern-.18em\Gamma$}}
 \def\IH{\relax{\rm I\kern-.18em H}}
 \def\II{\relax{\rm I\kern-.18em I}}
 \def\IK{\relax{\rm I\kern-.18em K}}
 \def\IP{\relax{\rm I\kern-.18em P}}

 \font\cmss=cmss10 \font\cmsss=cmss10 at 7pt
 \def\IR{\relax{\rm I\kern-.18em R}}

 %

 %
 %
 \def\eqnn#1{\xdef #1{(\secsym\the\meqno)}\writedef{#1\leftbracket#1}%
 \global\advance\meqno by1\wrlabeL#1}
 \def\eqna#1{\xdef #1##1{\hbox{$(\secsym\the\meqno##1)$}}
 \writedef{#1\numbersign1\leftbracket#1{\numbersign1}}%
 \global\advance\meqno by1\wrlabeL{#1$\{\}$}}
 \def\eqn#1#2{\xdef #1{(\secsym\the\meqno)}\writedef{#1\leftbracket#1}%
 \global\advance\meqno by1$$#2\eqno#1\eqlabeL#1$$}

 \lref\author{Name}
\lref\berg{E.~Bergshoeff,
 D.~S.~Berman, J.~P.~van der Schaar and P.~Sundell, ``A
 noncommutative M-theory five-brane,'' hep-th/0005026.}

\Title
 {\vbox{
 \baselineskip12pt
 \hbox{HUTP-01/A003}
 \hbox{hep-th/0101218}\hbox{}\hbox{}
}}
 {\vbox{
 \centerline{Brane/anti-Brane Systems and $U(N|M)$ Supergroup}
 }}
 \centerline{ 
Cumrun ${\rm Vafa}$}
 \bigskip\centerline{ Jefferson Physical Laboratory}
 \centerline{Harvard University}
\centerline{Cambridge, MA 02138, USA}
 \smallskip
 \vskip .3in \centerline{\bf Abstract}
{We show that in the context of topological
string theories  N branes and M anti-branes 
give rise to Chern-Simons gauge theory with the gauge supergroup
$U(N|M)$.  We also identify a deformation of the theory which
corresponds to brane/anti-brane annihilation.  Furthermore we show
that when $N=M$ all open string states are BRST trivial in the deformed theory.}
 \smallskip 
\Date{January 2001}

\newsec{Introduction}
Topological strings propagating on Calabi-Yau threefolds
provide a consistent vacuum of bosonic string theory
\ref\witten{E. Witten, ``On the structure
of the topological phase of two dimensional gravity,''
Nucl. Phys. {\bf B340} (1990) 281.}\ similar, and in some cases equivalent,
to non-critical vacua of the bosonic strings.  For example
the B-model topological string on the conifold is equivalent
to non-critical bosonic string propagating on a self-dual circle
\ref\ghov{D. Ghoshal and C. Vafa, ``c=1 string as the topological
theory of the conifold,'' Nucl. Phys. {\bf B45} (1995) 121.}.

One can also consider open string version of these theories
\ref\witcs{E. Witten, ``Chern-Simons theory as a string theory,''
hep-th/9207094.}\ which can be interpreted,
in modern terminology, as adding D-branes to the Calabi-Yau.
For example, if one considers $N$ D-branes wrapping a Lagrangian
3-manifold $L$ inside a Calabi-Yau, then the topological
A-model gives rise to an open string field theory in
the target space which is equivalent to ordinary Chern-Simons
theory on $L$.  In particular
if we consider $N$ D-branes on $L$ we obtain
$U(N)$ Chern-Simons theory on $L$ (similarly in the B-model
version one can
consider D-branes wrapping holomorphic cycles and one
ends up with a holomorphic version of Chern-Simons
theory as the open string field theory).

The aim of the present note is to generalize
these constructions to also include ``anti-D-branes''.
  We find that, with a suitable notion of 
what ``anti-D-branes'' are in the topological theory,
the theory involving $N$ D-branes and $M$ anti-D-branes
wrapping the three manifold $L$ in the context of A-type
topological strings,
leads to a Chern-Simons open string field theory
with the gauge groups being the supergroup $U(N|M)$ (and similarly
for the B-type case which leads to holomorphic Chern-Simons
theory with supergroup connection).  We also discuss the brane/anti-brane
annihilation by giving a vev to a  scalar field in the theory (which
is the topological analog of ``tachyon condensation'')
and show how the physical states of the open string theory are removed
upon this deformation. 

Certain aspects of brane/anti-brane systems in the context of topological
string theory and in particular its relation (in the
context of B-model topological strings) with
derived category has been pointed out in \ref\doug{M. Douglas,
``D-branes, categories and N=1 supersymmetry,'' hep-th/0011017.} .

\newsec{Review of Chern-Simons Theory as a String Theory}

Consider A-model topological strings propagating on Calabi-Yau
threefold, together with a Lagrangian submanifold $L$.  Putting
$N$ D-branes on $L$ we end up with the open topological string
sector involving $U(N)$ Chern-Simons gauge theory \witcs ,
$$S={ik\over 4\pi^2}\int_L Tr[{1\over 2} AdA +{1\over 3}A^3] $$
where the level
$ik$ (or $i(k+N)$ after quantum corrections)
of the Chern-Simons theory is identified with the inverse of the
string coupling constant.  In other words the open string worldsheets
in this context are exactly the `t Hooft diagrams of the $U(N)$
Chern-Simons theory.

  The only physical
string mode in the open string sector is the 1-form connection
$A$ and none of the oscillator modes of the open
string theory are physical.  In particular
the general open string Chern-Simons field theory coincides
with the ordinary Chern-Simons theory in this case.  Moreover
the BRST operator of the string $Q$ is identified with the
$d$ operator
$$Q\leftrightarrow d$$
The space of vacua is parameterized by flat connections $F=0$.
For each such vacuum, corresponding to connection $A_0$
the BRST cohomology gets deformed to the
covariant connection
$$Q\leftrightarrow d_{A_0}=d+[A_0,.]$$
which can be readily seen by expanding the action near
the new vacuum up to quadratic terms in the fluctuation field $\delta A$.
For each such vacuum
the physical states of the theory are in one to one correspondence
with $Q$ cohomology, which is the same as the tangent space to the
moduli space of flat connections at $A=A_0$. Note that if there
is no continuous family of flat connections, as would be the case
if $\pi_1(L)$ is empty, there would be no physical state
in the open string sector.  This does not necessarily imply
that open string sector gives zero partition function.  For example
if we take $L=S^3$ then there are no (local) physical states  as the
fundamental group of $S^3$ is trivial, but nevertheless the partition 
function of $U(N)$ Chern-Simons is highly non-trivial (in other
words the worldsheet diagrams with boundaries do not give zero).
There are in addition non-local observables
for Chern-Simons theory, such as Wilson lines along
links.\foot{ It might be interesting to study
the analog of Wilson loop observables for ordinary open string field theory
which is also abstractly a Chern-Simons theory
(in Witten's formulation).}

The open string field theory comes in general equipped with the ghosts
which provide the BRST fields for the
open string field theory.  This is also the case in this particular
example. The ghosts of the Chern-Simons gauge theory
in this case correspond to enlarging $A$ to be a combination of
all forms \witcs \ref\singerax{S. Axelrod and I.M. Singer,
``Chern-Simons perturbation theory--II,'' J. Diff. Geom. {\bf 39}
(1994) 173.}, with even forms being fermionic and odd forms being bosonic.
The ghost number of the string field theory coincides with the
degree of the form. The usual physical field corresponds to ghost number 1,
which is the one-form connection. However in topological string theory it is
natural to consider all ghost number fields (including
their zero modes)
 on equal footing and consider
the extended theory.  In particular the bosonic
fields of this extended theory can in general have a bigger
moduli space than the original theory as is
well known in the context
of closed topological strings (where it is called the extended moduli
space).  We will consider this extended theory also in the
context of open strings and this will play an important role
when we discuss brane/anti-brane annihilation below. 

One can also consider the B-model topological strings and in this
case (if we consider N wrapped D6 branes) we obtain a holomorphic
version of Chern-Simons theory, whose basic field is a $(0,1)$-form
connection ${\overline A}$ of a holomorphic
bundle, with the action
$$S={ik\over 4\pi^2}\int \Omega \wedge Tr[{1\over 2}{\overline A}{\overline 
\partial} {\overline A}+{1\over 3}{\overline A}^3]$$
where $\Omega $ is the holomorphic 3-form on the Calabi-Yau.
Here $Q={\overline \partial}+[{\overline A},.]$.
Similarly on lower even dimensional branes one obtains
the natural reduction of the above action to the corresponding dimension.
 This extended
theory (i.e. including all degree forms as part of ${\overline A}$)
has also
been considered recently in \ref\jsp{J.-S. Park, ``Topological Open p-Branes,''
hep-th/0012141.}.

\newsec{Topological Anti-D-branes}

We should decide what
in the topological theory should distinguish
 branes from anti-branes.  A natural choice, which
we will adopt, is
 the following:  The anti-branes should carry the opposite
number of branes compared to branes.  In the topological
theory the only signature of dealing with a number of branes is the
Chan-Paton factor $N$ that one associates to each 
worldsheet boundary ending on them.
Thus if we have $M$ anti-branes, it is natural to associate
a factor of $-M$ for each such hole.  Put differently, for each
anti-brane we put an extra minus sign for each worldsheet 
hole ending on it.  This is our definition of topological
anti-D-branes.  Note that if we have only anti-branes
wrapping some Lagrangian 3-cycle, the net effect of this
on the partition function is the same as weighing
worldsheet diagrams with odd number of holes with
an extra minus sign.  This in turn can be viewed
as replacing the string coupling by minus itself,
or replacing $ik\rightarrow -ik$ (or in the quantum
corrected version $i(k+N)\rightarrow -i(k+N)$), which is
the same as complex conjugation of the partition function.
This is consistent
with what one might naturally expect for anti-branes.

In the next section, we use the above definition
of anti-branes and consider a situation where we have
both branes and anti-branes and write down the effective
open string field theory.

\newsec{ Brane/anti-Brane Systems and Chern-Simons Gauge Theory
with Supergroup $U(N|M)$}

Consider $N$ branes and $M$ anti-branes wrapped around
a Lagrangian 3-cycle $L$.  We would like to know what is
the open string field theory for this theory.  The physical
state of the open string sector can be deduced, as usual,
by considering the annulus diagram.  In this case there
are four such diagrams; One coming from boundaries ending
on branes carrying a factor of $N^2$--the corresponding
physical state (at ghost number one) is naturally
interpretable as the one-form connection in a $U(N)$
gauge group.  The other diagram comes from
both boundaries ending on an anti-Brane.  This gives
a factor of $(-M)(-M)=+M^2$, whose physical
field can be
interpreted as a one form connection for the gauge group
$U(M)$.  We also have two more diagrams ending
on opposite type of branes, each giving a factor
$N(-M)=-NM$.  The corresponding physical states
can be interpreted as 1-form gauge fields which
are in the representation 
$$(N,{\overline M})\oplus
({\overline N},M)$$
However, the extra minus sign in this annulus
diagram implies that they are to be viewed as fermionic
fields.  In fact these fields naturally assemble
themselves into the connection 1-form ${\cal A}$
for a supergroup $U(N|M)$.  It is also similarly
straightforward to derive the open string field theory.
It is the Chern-Simons gauge theory for the supergroup
$U(N|M)$:
$$S={ik\over 4\pi ^2}\int_L Tr[{1\over 2}{\cal A}d
{\cal A} +{1\over 3}{\cal
A}^3]$$
where the notion of trace in the above action is in terms of
the natural trace for the supergroup $U(N|M)$.  The appearance
of `superconnection' is also natural from the viewpoint
of brane/anti-brane systems in ordinary superstring theory
\ref\witk{E. Witten, ``D-branes and  K-theory,'' JHEP9812 
(1998) 019.}\ref\krla{P. Kraus and F. Larsen,
``Boundary string field theory of the DDbar system,''
hep-th/0012198.}\ref\jap{T. Takayanagi, S. Terashima
and T. Uesugi,
``Brane-Antibrane action from boundary string field theory,''
hep-th/0012210.}\ref\ali{M. Alishahiha, H. Ita and Y. Oz,
``On superconnections and the tachyon effective action,'' hep-th/0012222.}.
However, in the superstring context the off-diagonal elements that appear
in the superconnection are not connections (i.e. are not
1-forms), but are replaced by tachyons.  Thus the superconnection
in the usual superstring context, unlike here, cannot be viewed
as a connection for a supergroup\foot{It would be interesting
to see if there is any sense in which a gauge supergroup
also exists in the superstring context.}.  In the next section
we will discuss, in the topological context, how the analog
of tachyon
field arises.

Let us note that the partition function of the $U(N|M)$
Chern-Simons theory
(expanded near the trivial connection) is the same
as that of $U(N-M)$.  This follows from the simple
observation \ref\bvw{N. Berkovits, C. Vafa and E. Witten,
``Conformal field theory of AdS background with Ramond-Ramond
flux,'' JHEP9903 (1999) 018.}\
that in the `t Hooft expansion of $U(N|M)$
gauge theory, the $N,M$ dependence for each `t Hooft
organization of Feynman diagram
arises from boundary Chan-Paton factors, and 
 for each boundary we obtain a
supertrace
over the fundamental representation of $U(N|M)$
which gives $(N-M)$.
 This is of course manifest from our definition
of what anti-branes are.
  Thus the amplitudes
for the Chern-Simons theory with supergroup $U(N|M)$
and that with gauge group $U(N-M)$ coincide,
as one might expect for a theory with $N-M$ net
number for branes.  This
again lends further support for the above proposal
of what anti-branes are.  In fact it has already been
noted that also certain knot invariants for $U(N|M)$
theory are the same as that of $U(N-M)$ theory \ref\supre{J.H.
Horne, ``Skein relations and Wilson loops in Chern-Simons gauge theory,''
Nucl. Phys. {\bf B334} (1990) 669.}\ref\anot{M.
Bourdeau, E.J. Mlawer, H. Riggs and
H.J. Schnitzer, ``The quasirational fusion structure of
$SU(M|N)$ Chern-Simons and W-Z-W theories,'' Nucl. Phys. {\bf B372}
(1992) 303.}, though some subtleties arise when $N=M$ \ref\salr{
L. Rozansky and H. Saleur, ``Reidemeister torsion, the Alexander
polynomial and $U(1,1)$ Chern-Simons theory,''
J. Geom. Phys. {\bf 13} (1994) 105.}.  
Apriori the above argument about `t Hooft
diagram only implies the equality of partition functions
and it is perfectly consistent with existence of
certain observables of $U(N|M)$
theory which are not present in the $U(N-M)$ theory.
Also if we expand about certain connections where the
connection in $U(M)\subset U(N)$  are not equivalent, apriori there
is no reason for the equivalence even of the partition function,
with that of $U(N-M)$.  This would then correspond
to a situation where the branes and anti-branes cannot
perfectly cancel one another even if the ``tachyon mode''
is turned on.

Note that the above discussion naturally generalizes to the
case of topological B-models where one ends up getting
holomorphic Chern-Simons theory with supergroup $U(N|M)$.
Also one can formulate the reduction of this to lower dimensional
holomorphic branes, and for example, study the topological
version of the D0 branes dissolving in D2 branes.

\newsec{Topological Analog of Brane/anti-Brane Annihilation}

It is natural to ask if there is any topological
analog of tachyon condensation and brane/anti-brane
annihilation along the lines proposed by Sen
\ref\sen{A. Sen, ``Tachyon condensation on the
brane antibrane system,'' JHEP 9808 (1998) 012.}.  At first the answer
might appear to be in the negative, because the field
which would have played the role of the tachyon is fermionic
1-form and it cannot take a vev.  However upon closer inspection,
as we will now argue the ``tachyon field'' has a more subtle
presence in the Chern-Simons theory.  We will identify a field
whose vev plays the role of the vev for the tachyon field.
However, there is no potential for the scalar field and in particular
it is not tachyonic in the usual sense.

For the case at hand, namely the Chern-Simons
gauge theory with supergroup $U(N|M)$,
just as in the case of ordinary gauge group $U(N)$, the
open string field theory including the ghosts
needed for BRST gauge fixing, leads to replacing
${\cal A}$ by an arbitrary degree form, with alternating statistics.
For the diagonal blocks of $U(N|M)$ the even forms are fermionic
and odd forms are bosonic,
whereas for off-diagonal block the odd forms of ${\cal A}$
 are fermionic and the even forms are bosonic.
  We will consider
this extended theory, and treat
all components of the extended ${\cal A}$ field on the same footing.
In particular there are {\it scalar} bosonic ghost
fields in the off-diagonal blocks.  These fields will play
the role of the ``topological tachyon fields''. 
In fact we will now demonstrate
that turning them on is allowed (i.e. can 
lead to new classical solutions) and in case of $N=M$ and
when the branes have the same gauge field configuration,
trivializes the string observables.

Even though we have scalar analogs of tachyon fields, there are
no scalar potentials for them.  However we can still ask
whether there are any other vacua for Chern-Simons
theory, by turning them
on\foot{Strictly speaking this makes sense in non-compact situation,
where constant mode fluctuation is not allowed.  We can assume
we are in the non-compact case,
for example when $L=R^3$ in a non-compact CY such
as ${\bf C}^3$--but
one can also make sense of it in the compact case as field configurations
which satisfy the equations of motion, by first integrating
over the non-zero modes in the path-integral.}.
We would like to identify
the Chern-Simons analog of other critical points of tachyon potential.
This means finding other classical solutions for
the open string field theory.
To satisfy the equations of motion, is equivalent
to finding a vacuum which has a BRST operator $Q$ which
squares to zero: $Q^2=0$.  As discussed before, this is equivalent
to considering gauge field configurations where the curvature
is zero, i.e.,
$$Q^2=0\rightarrow d_{\cal A}^2=(d+[{\cal A},.])^2=0
\rightarrow {\cal F}=0$$
Note that ${\cal F}$ is the curvature of the connection
for the supergroup $U(N|M)$.  Moreover, if we allow the
bosonic ghosts also to take vev, this is equivalent to allowing
the connection to have vevs for the odd forms in the diagonal
blocks, and vevs for the even forms in the off-diagonal blocks.

Let us give an example of how this works: 
Consider two gauge field configurations for the bosonic
parts of the supergroup $U(N)\times U(M)$, characterized
by connections $A=A_N+A_M$.  Suppose there is a covariantly
constant field $\phi$ in the bifundamental $(N,{\overline M})$, i.e.,
$$d_A\phi =d\phi +A_N \phi -\phi A_M =0$$
If we give a vev to this off diagonal component of $U(N|M)$
at degree 0, i.e. the ``bosonic ghost field'', then
we have another vacuum for the Chern-Simons theory with the 
gauge supergroup.
In particular expanding to quadratic order, we can identify
the new BRST operator
$$Q=d_{\cal A}=d_A +[\phi ,.]$$
where written in this form we view $\phi$ as an upper block off-diagonal
matrix valued element of $U(N|M)$ adjoint representation.   Note that
$Q^2=0$ and we can thus view this as a new vacuum of the theory.
This is also consistent with the viewpoint advocated in 
\doug\ for cohomological aspects of brane/anti-brane annihilation
and its relation to derived category.

The physical
states in the new vacuum will in general be different, because
now we have a different cohomology problem defined by $Q$. Let us specialize
to the following case:  Suppose
 we have equal number of branes and anti-branes $N=M$ and assume
we have the same flat $U(N)$ gauge connection on both types of branes. 
Then we have a natural
choice for $\phi$, namely the identity $N\times N$ matrix.  
We will identify giving a vev
to this field as ``brane/anti-brane annihilation'', which is the analog
of tachyon condensation in this topological setup.  As a test
of this idea we show that the open string field theory BRST operator
$Q$ for this new vacuum after giving
a vev to $\phi$ trivializes the cohomology elements of the
old $Q$. In particular suppose we originally have a number
of deformations for the flat bundle for both $U(N)$'s, which
by our assumption they are the same.  Restricted to each of these modes
the new $Q$ operator will reduce to $Q=[\phi, .]$.  It is easy to see
that with respect to an upper block off-diagonal 
identity matrix $\phi$ all the cohomolgy
disappears:  We can view the $(N|N)$ adjoint as consisting of
four natural $(N\times N)$ blocks. We now show that the new
$Q$ pairs up these four blocks.
The lower off diagonal block
is not $Q$ invariant (as it does not commute with $\phi$).  It instead
gives the equal sum of the two diagonal blocks.  On the other hand
$Q$ acting on
the opposite combination of the two diagonal blocks will give 
all upper block off-diagonal combinations.  Thus
all the cohomologically non-trivial elements are now paired
up and we have no
non-trivial
cohomology left, as expected for a theory which is to be
identified with a trivial theory.  Note that we have already
explained why the partition function will be trivial, when
the brane/anti-brane bundles are expanded near the trivial connection.
That argument can be extented to the case where the connections in the
two $U(N)$'s are the same. But consider a situation where
$L$ has a non-trivial $\pi_1$.  In that case we can consider
inequivalent configurations for the two gauge groups, and
from the perspective of `t Hooft diagram, there is no reason
why such configurations should give vanishing contribution
to the partition function.  It is amusing that precisely for
these cases there is no
covariantly constant identity matrix
$\phi$ (i.e. precisely for the case where we do not expect
complete annihilation there is also no reason for the partition function
to vanish).

\newsec{Potential Applications to Superstring Compactifications}

As discussed in \ref\vaau{C. Vafa, ``Superstrings and topological
strings at large $N$,'' hep-th/0008142.}, topological strings with
branes wrapped over Lagrangian submanifolds compute
certain terms in the effective theory of type IIA strings compactified
on Calabi-Yau manifolds with partially wrapped D6 branes, filling
the spacetime.  Also, as discussed there, in the context of
compact Calabi-Yau we have to consider equal number of branes
and anti-branes, and for some such cases, a large $N$ dual was
proposed.  Given that we have
defined what topological superstrings are in the context of branes
and anti-branes, it is natural to ask if there are similar
terms in the type IIA superstring context with wrapped branes and
anti-branes that they compute. This of course would sound
very remarkable given that this would correspond to some exact
computations in a non-supersymmetric background. However,
this may not be as surprising.  For example, the net
number of branes is a computation which one expects
to be able to do exactly in the non-supersymmetric cases.
Also superpotential terms are another example of what branes
and anti-branes give rise to that should be exactly computable
(similar to turning on RR fluxes in Calabi-Yau, which for a generic
moduli is not supersymmetric, but give rise to computable
superpotential terms).
The terms that the topological superstrings compute
in the cases including branes and anti-branes must
be generalizations of such terms and it would be very interesting
to precisely identify them.  

\vglue 1cm

I would like to thank the hospitality of the string theory
group at Rutgers University.  In particular I have greatly benefitted
from discussions with M. Douglas, M. Marino and G. Moore.
I would also like to thank N. Berkovits, K. Hori, S. Katz and
I.M. Singer for valuable discussions.

This research is supported in part by NSF grants PHY-9802709
and DMS 9709694.
\listrefs

\end